\documentclass[]{aipproc}

\layoutstyle{6x9}

\SetInternalRegister\hbadness{8000} %

\newcommand\doingARLO[2][]{%
  \ifx\mmref\undefined #1\else #2\fi
}

\def\NIMA{{\em Nucl. Instrum. Methods} A}
\def\NIMB{{\em Nucl. Instrum. Methods} B}
\def\NPB{{\em Nucl. Phys.} B}
\def\PLB{{\em Phys. Lett.}  B}
\def\ARNPS{{\em Ann. Rev. Nucl. Part. Sci.}}
\def\PRL{\em Phys. Rev. Lett.}
\def\PRD{{\em Phys. Rev.} D}
\def\EPJC{{\em Eur. Phys.} C}

\def\be{\begin{equation}}
\def\ee{\end{equation}}
\def\bea{\begin{eqnarray}}
\def\eea{\end{eqnarray}}

\begin{document}

\title
{Recent results of the STAR high-energy polarized proton-proton program at RHIC at BNL} 

\author{Bernd Surrow (For the STAR Collaboration)}{
  address={Massachusetts Institute of Technology \\ 77 Massachusetts Avenue, Cambridge, MA 02139, USA},
  email={surrow@mit.edu}
  thanks={}
}

\iftrue

\fi

\copyrightyear  {2001}

\begin{abstract}
The STAR experiment at the Relativistic Heavy-Ion Collider (RHIC) at Brookhaven National
Laboratory (BNL) is carrying out a spin physics program colliding transverse
or longitudinal polarized proton beams at $\sqrt{s}=200-500\,$GeV 
to gain a deeper insight into the spin structure and
dynamics of the proton. These studies provide fundamental tests of Quantum Chromodynamics (QCD).

One of the main objectives of the STAR spin physics program is the determination of
the polarized gluon distribution function through a measurement of the longitudinal double-spin 
asymmetry, $A_{LL}$, for various processes. 
Recent results will be shown on the measurement of $A_{LL}$
for inclusive jet production, neutral pion production and charged pion production at $\sqrt{s}=200\,$GeV.
In addition to these measurements involving longitudinal polarized proton beams, the STAR
collaboration has performed several important measurements employing transverse polarized proton beams.
New results on the measurement of the transverse single-spin asymmetry, $A_{N}$, for forward neutral
pion production and the first measurement of $A_{N}$ for mid-rapidity di-jet production will be discussed.
\end{abstract}

\date{\today}

\maketitle

\section{Introduction}
  
The core goal of the RHIC spin program is to obtain a deeper understanding of the spin structure and 
dynamics of the proton in polarized proton-proton collisions \cite{ref_bunce}. 
Shedding light on the proton spin puzzle by providing insight on how the intrinsic spin of the proton
is distributed among its underlying constituents of quarks, anti-quarks and gluons is an important aspect
of the program. Determination of the parton orbital angular momentum contributions and gluon helicity distribution are 
essential for a complete understanding of the proton spin.

\begin{figure}[h]
\setlength{\unitlength}{\textwidth}
\begin{picture} (1.0,0.375)
\put (0.05,0.0){\mbox{\includegraphics*[width=55mm]{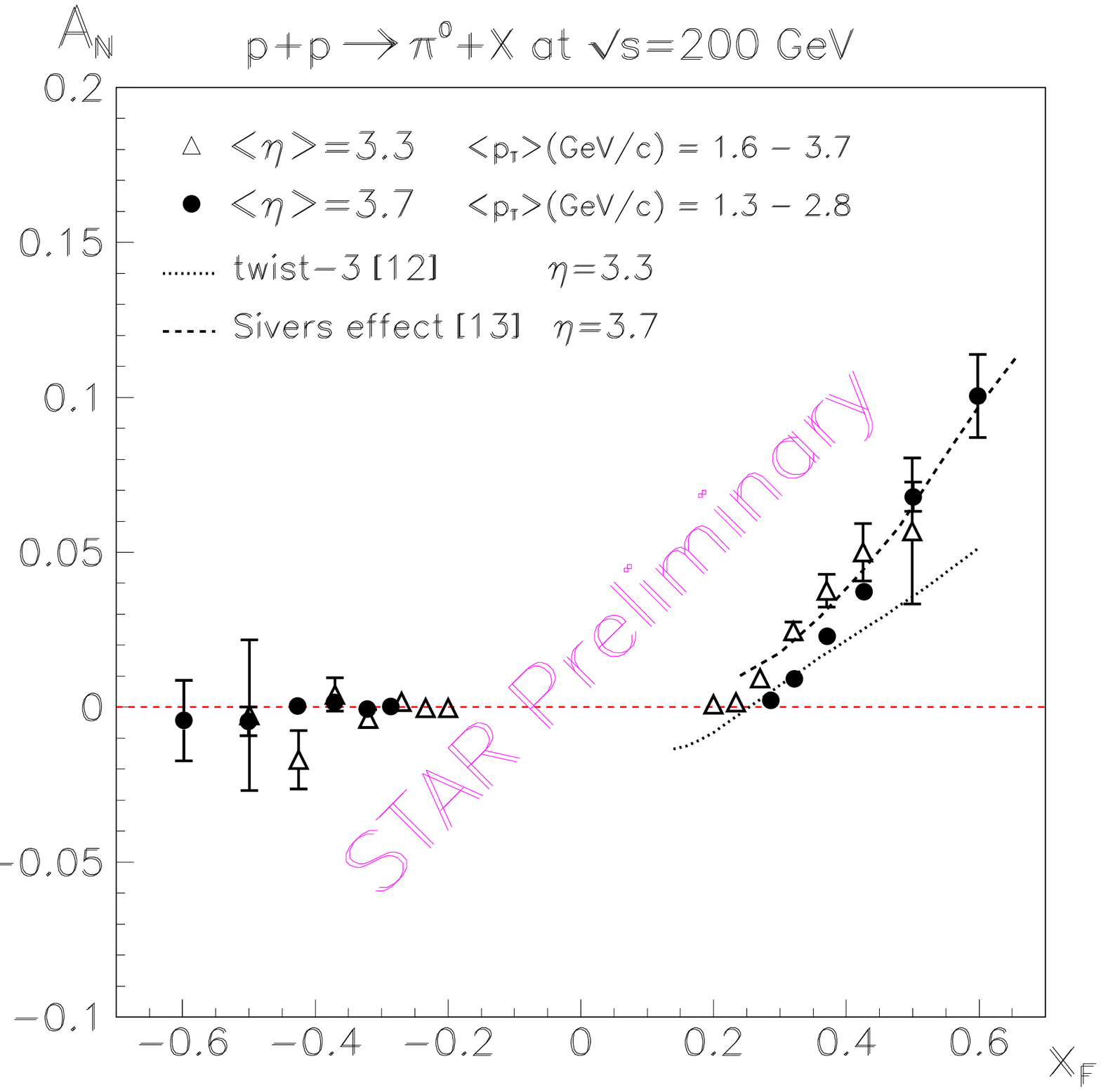}}}
\put (0.56,0.0){\mbox{\includegraphics*[width=55mm]{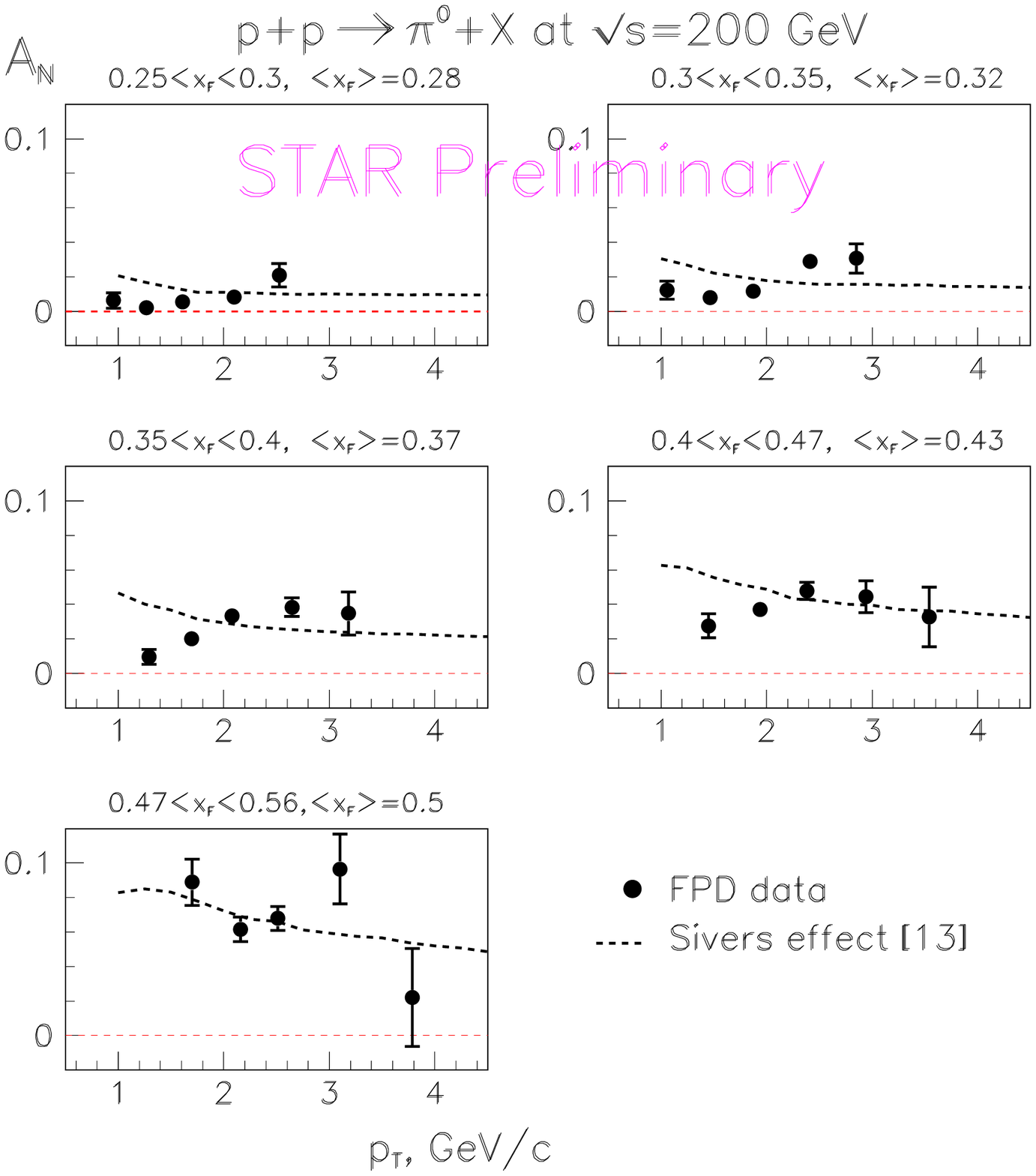}}}
\end{picture}
\caption{\it (left) $A_{N}$ as a function of $x_{F}$ for $p+p\rightarrow \pi^{0}+X$ from Run 6. The error bars represent statistical
errors only. Systematic uncertainties are smaller than statistical uncertainties. The dotted line represents the 
result of a twist-3 calculation \cite{ref_kouvaris}, whereas the dashed line is the result of a calculation based on the Sivers
effect \cite{ref_alesio}. (right) $A_{N}$ as a function of $p_{T}$ for $p+p\rightarrow \pi^{0}+X$ from Run 6 in bins of $x_{F}$ in comparison to a calculation based on the Sivers
effect \cite{ref_alesio}.} 
\label{fig:an_2006_xf_pt}
\end{figure}

The RHIC facility is the first 
polarized collider, providing collisions of transverse or longitudinal polarized proton beams at a 
center-of-mass energy of $\sqrt{s}=200\,$GeV and in the future of $\sqrt{s}=500\,$GeV. The STAR spin physics program 
has profited enormously from the steady improvement and development of the RHIC polarized proton-proton collider facility in 
terms of polarization and luminosity. The performance of the most recent run in 2006 (Run 6) is very encouraging 
with an average polarization of $60\%$ and a delivered luminosity per day of approximately $1\,$pb$^{-1}$ at $\sqrt{s}=200\,$GeV. This is 
to be compared with the design performance of $70\%$ in beam polarization and a daily delivered 
luminosity of approximately $3\,$pb$^{-1}$ at $\sqrt{s}=200\,$GeV. Several anticipated improvements along with the required 
subsequent development of the RHIC facility are expected to yield the 
design performance at $\sqrt{s}=200\,$GeV and $\sqrt{s}=500\,$GeV \cite{ref_cad}. An 
important step for the STAR spin physics program was the first transverse 
single-spin asymmetry measurement in forward neutral pion production from the first polarized proton-proton 
run in 2002 (Run 2) \cite{ref_fpd_an} and the first longitudinal double-spin asymmetry measurement sensitive to the gluon 
polarization in mid-rapidity inclusive jet production from Run 3 / 4 \cite{ref_all_0304}.    

The following two sections will highlight recent results \cite{ref_spin2006_jan, ref_spin2006_adam, ref_spin2006_larissa, ref_spin2006_dave, ref_spin2006_frank, ref_spin2006_jason, ref_spin2006_qinghua} of the STAR transverse and longitudinal spin 
physics program. 

\section{STAR transverse spin program - Recent results}  
 
The study of transverse spin effects, both theoretically and experimentally, has received 
a lot of attention in recent years. The ultimate goal of this effort is to extract transversity 
distribution functions and determine possible orbital angular momentum contributions of quarks, anti-quarks
and gluons to the proton spin \cite{ref_A.Ogawa}. 

The first measurement of a transverse single-spin asymmetry, $A_{N}$, at $\sqrt{s}=200\,$GeV for forward
neutral pion production at $3.3<\eta<4.1$ by the STAR collaboration \cite{ref_fpd_an} was found to increase 
with $x_{F}$ similar in magnitude to the measurement of 
$A_{N}$ performed by the E704 experiment at $\sqrt{s}=20\,$GeV \cite{ref_E704}. The STAR collaboration has performed cross-section
measurements for forward neutral pion production at $\langle \eta \rangle=3.3$, $3.8$ and $4.0$, which are found
to be in good agreement with next-to-leading order (NLO) perturbative QCD calculations \cite{ref_fpd_cross}. This provides an 
important basis for the interpretation of the sizable measured asymmetries at forward rapidity. Several
models beyond a conventional collinear perturbative QCD approach are able to account for the measured asymmetries. These models 
are based on three main correlation concepts: A correlation of the parton intrinsic transverse momentum $k_{T}$ and the proton spin in the initial state 
(Sivers effect) \cite{ref_sivers}, a correlation between the quark spin and the hadron $k_{T}$ in the final state (Collins effect) \cite{ref_collins} and 
a higher twist correlation in the initial and final state \cite{ref_qui_sterman, ref_efremov}. The STAR collaboration has
performed an upgrade of their Forward Pion Detector (FPD) for Run 6 as a prototype for the Forward Meson Spectrometer (FMS).
The main goal is to disentangle different mechanisms responsible for the observed large transverse single-spin asymmetries
in the forward direction. 
New results of forward neutral pion production at higher precision are shown in Figure \ref{fig:an_2006_xf_pt} \cite{ref_spin2006_larissa}. 
These results
are in good agreement with previous results. The measured asymmetry $A_{N}$ is found to be consistent with zero at
negative $x_{F}$ and grows from $0$ at $x_{F}\simeq 0.2-0.3$ up to $0.1$ at $x_{F}\simeq 0.6$. Figure \ref{fig:an_2006_xf_pt} shows also
the result of two calculations at different $\langle \eta \rangle$ values \cite{ref_kouvaris, ref_alesio}. 
The measured asymmetries are precise enough
to allow for a quantitative comparison with theory predictions. The extended kinematic region in $x_{F}$ and $p_{T}$ and
the increase in the available data sample from Run 6 allowed for the first time for a mapping of $A_{N}$ in $x_{F}$ and
$p_{T}$. Figure \ref{fig:an_2006_xf_pt} shows $A_{N}$ as a function of $p_{T}$ in fine bins of $x_{F}$. 
The current measured result of $A_{N}$ do not support a decreasing behavior of $A_{N}$ in $p_{T}$ in all $x_{F}$ bins as 
expected by various theoretical models.  

\begin{figure}[t]
\centering
\includegraphics*[width=85mm]{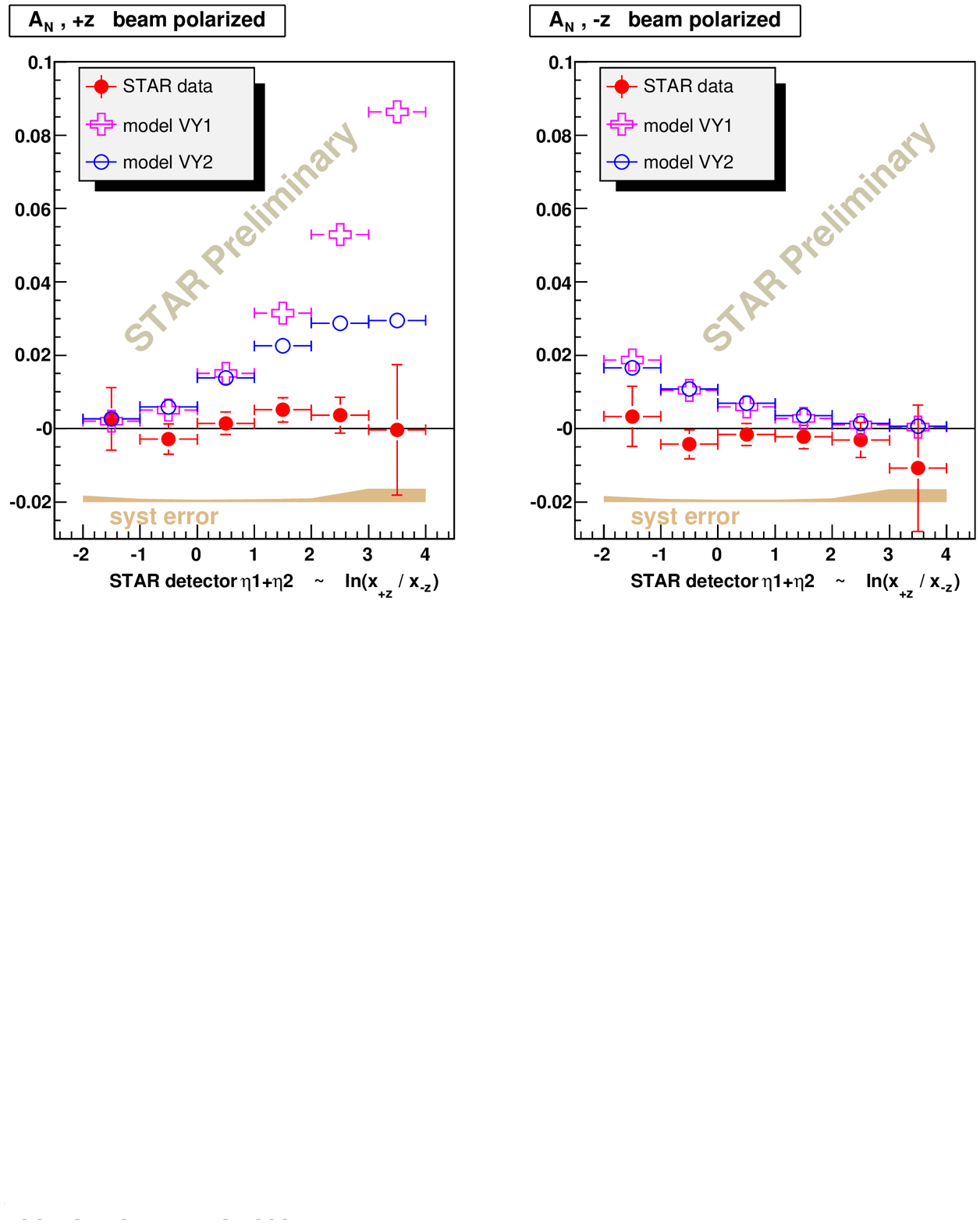}
\caption{\it Transverse single-spin asymmetries for $A_{N}^{+z}$ (left) and $A_{N}^{-z}$ (right) as a function of
$\eta_{1}+\eta_{2}$ for di-jet production at $\sqrt{s}=200\,$GeV from Run 6 in comparison to theoretical calculations \cite{ref_v_y, ref_werner}.}
\label{fig:an_2006_dijets}
\end{figure}

A possible manifestation of orbital angular momentum effects could be realized through a 
non-zero Sivers type correlation of the parton intrinsic transverse momentum $k_{T}$ and the proton spin in the initial state.
The HERMES collaboration has recently reported a non-zero Sivers function in semi-inclusive DIS for $\pi^{+}$ production \cite{ref_an_hermes}. 
The Sivers function for $\pi^{-}$ production has been found to be consistent with zero. This led to the interpretation that the Sivers functions
are opposite in sign and different in magnitude for u compared to d quarks. The study of Sivers type correlations in polarized
proton-proton collisions has been considered in \cite{ref_bv}. The theoretical expectation is that the Sivers effect would be reflected in a spin-dependent
side-ways boost to the di-jet opening angle. A measurement of the correlation between the spin direction and di-jet bi-sector
direction would be needed. A dedicated Level 2 di-jet trigger, utilizing the STAR 
Barrel and Endcap Electromagnetic Calorimeters (BEMC and EEMC), was implemented during Run 6. 

Figure \ref{fig:an_2006_dijets} shows the measured transverse single-spin asymmetry, $A_{N}$, for di-jet production as function
of $\eta_{1}+\eta_{2}$ for both polarized beams in the $\pm z$ direction \cite{ref_spin2006_jan}. 
The region of large $\eta_{1}+\eta_{2}$ values emphasizes quark (gluon) Sivers effects for $+z$ ($-z$). In both cases, the measured transverse single-spin
asymmetries are found to be consistent with zero. The measured results are compared to predictions based on u and d
quark Sivers functions extracted from measured HERMES semi-inclusive DIS data, which assume a zero gluon-type Sivers
function \cite{ref_v_y, ref_werner}. The calculations were carried out using only gauge-link factors for initial-state interactions, as expected
for Drell-Yan production. Further modifications are required for di-jet production taking into account final-state
effects. Theoretical work on the proper treatment of these final-state effects is ongoing.

Further theoretical work is needed to account for both the large transverse single-spin asymmetries for
forward neutral pion production and the recent di-jet result from STAR in polarized proton-proton collisions, as well as the results obtained
in semi-inclusive DIS. The STAR Forward Meson Spectrometer, which is expected to
be operational in the 2007 RHIC run, will provide a full azimuthal coverage at $2.5<\eta<4.0$. This detector system will allow STAR
to isolate Sivers from Collins effects. It will extend the jet, photon and neutral pion acceptance of the STAR experiment,
which will be also of considerable interest for the longitudinal spin program.

\begin{figure}[t]
\begin{minipage}[t]{5cm}
\includegraphics[height=4.5cm,width=10.0cm]{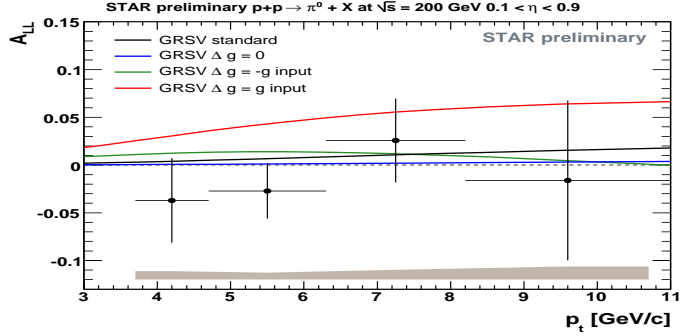}
\end{minipage}
\hfill
\begin{minipage}[t]{5.0cm}
\parbox[b]{5cm}{\caption{\it Longitudinal double-spin asymmetry $A_{LL}$ for neutral pion production at $\sqrt{s}=200\,$GeV as a function of $p_{T}$ ($0.1 < \eta < 0.9$) for Run 5 in comparison to several gluon polarization scenarios.}}
\end{minipage}
\hfill
\label{fig:all_2005_pions}
\end{figure}

\section{STAR longitudinal spin program - Recent results} 

The longitudinal STAR spin physics program profits enormously from the unique capabilities of 
the STAR experiment for large acceptance jet production, identified hadron production and 
photon production \cite{ref_star}. 
The measurement of the gluon polarization through inclusive measurements such as 
jet production and $\pi^{0}$ production has been so far the prime focus of the physics analysis program
of the Run 3/4 and Run 5 data samples. The sensitivity 
of these inclusive measurements to the underlying gluon polarization in high-energy polarized
proton-proton collisions has been discussed in detail in \cite{ref_grsv}.
Inclusive hadron production and jet production are strongly affected by the relative contributions
from quark-quark, quark-gluon and gluon-gluon subprocesses. The low $p_{T}$ region is dominated by gluon-gluon
scattering, while at high $p_{T}$ the quark-gluon contribution starts to become important. As a result, the sign
of $A_{LL}$ in this high $p_{T}$ region indicates the sign of the gluon polarization. The fact that the inclusive
photon channel is dominated by quark-gluon scattering results in a strong sensitivity to the underlying gluon polarization,
despite the small production cross-section.
Throughout the following discussion, four gluon polarization scenarios have been used as input to NLO perturbative QCD
calculations of $A_{LL}$.
The GRSV standard case refers to the best global analysis fit to polarized DIS data \cite{ref_grsv_global_fit}. The case 
for a vanishing gluon polarization (GRSV-ZERO) and the case of a maximally positive (GRSV-MAX) or negative (GRSV-MIN) 
gluon polarization have been also considered. 
 
The first longitudinal double-spin asymmetry measurement for inclusive 
jet production and the associated inclusive jet cross-section measurement at mid-rapidity has 
recently been published \cite{ref_all_0304}. The measured asymmetries are consistent with NLO perturbative QCD calculations based on DIS polarized 
lepton-nucleon data, and disfavor a large positive value of the gluon polarization in the proton.
In addition, the STAR collaboration has recently released a preliminary result on the inclusive neutral pion 
production cross-section at mid-rapidity \cite{ref_frank_cipanp}. The (unpolarized) cross-section measurements of inclusive
jet and neutral pion production support the asymmetry measurements at RHIC by validating the applicability of perturbative QCD. 

\begin{figure}[t]
\setlength{\unitlength}{\textwidth}
\begin{picture} (1.0,0.275)
\put (0.1,0.0){\mbox{\includegraphics*[width=55mm]{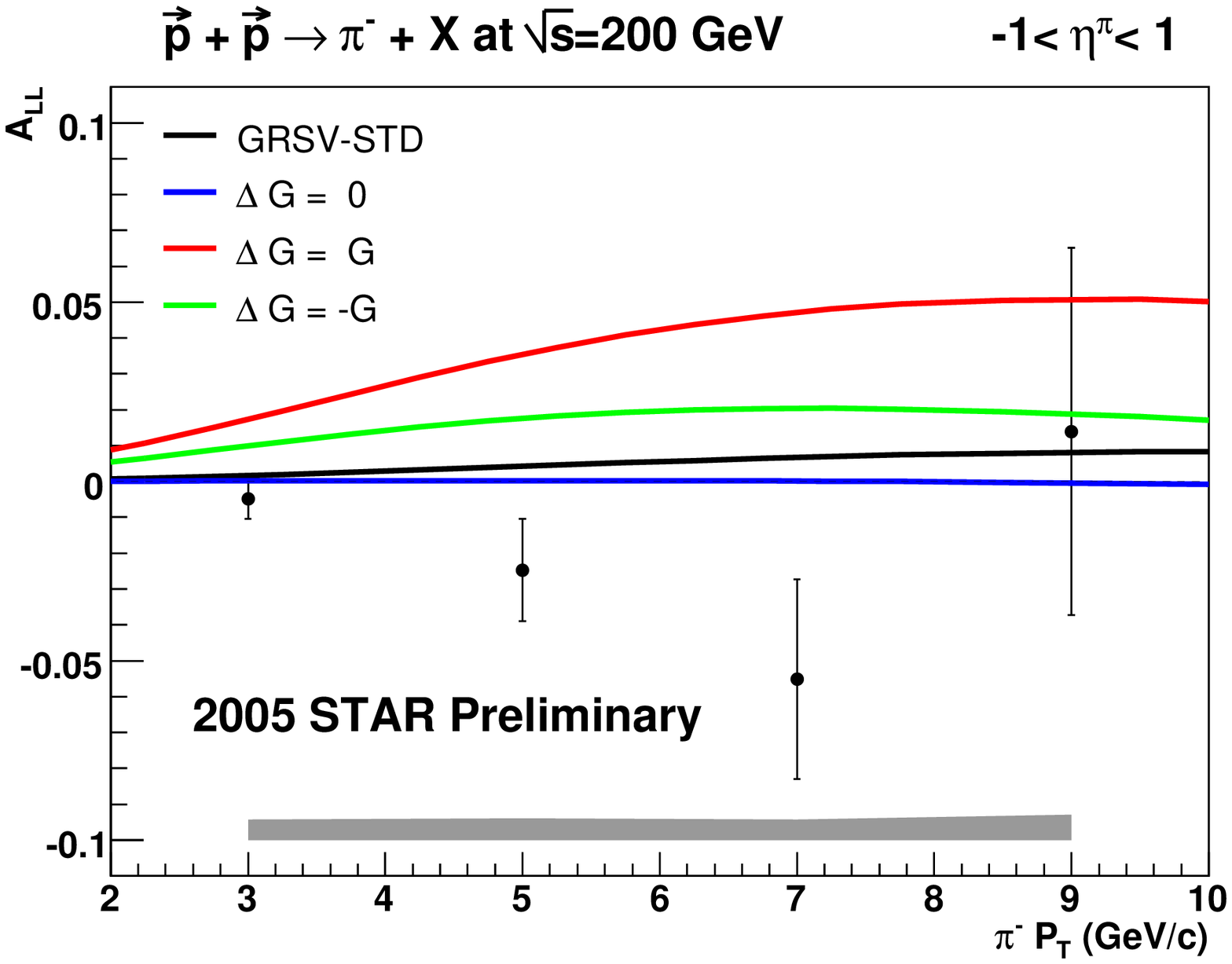}}}
\put (0.5,0.0){\mbox{\includegraphics*[width=55mm]{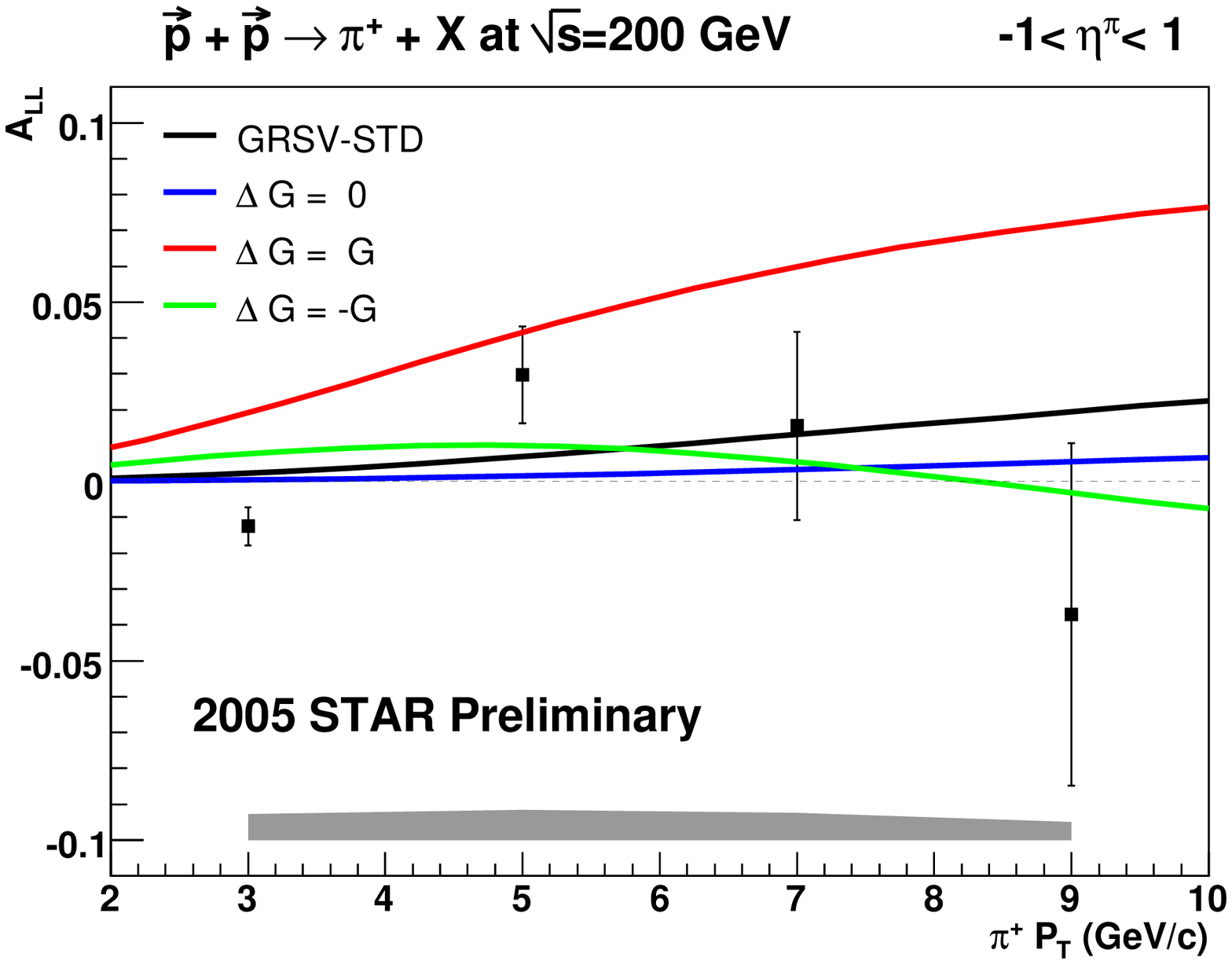}}}
\end{picture}
\caption{\it Longitudinal double-spin asymmetry $A_{LL}$ for charged pion production at $\sqrt{s}=200\,$GeV as a function of $p_{T}$ 
($-1 < \eta < 1$) for Run 5 in comparison to several gluon polarization scenarios.}
\label{fig:all_2005_pi-_pi+}
\end{figure}

The following measurements of $A_{LL}$ from Run 5 at $\sqrt{s}=200\,$GeV 
\cite{ref_spin2006_adam, ref_spin2006_dave, ref_spin2006_frank, ref_spin2006_jason} 
for identified hadron production and inclusive jet production 
are based on an average beam polarization of approximately $50\%$ and 
and a data sample of approximately $3\,$pb$^{-1}$. The overall normalization uncertainty 
due to conservative error estimates on the preliminary polarization values amounts to $\sim 40\%$. 
All $A_{LL}$ analyses presented below make use of the STAR BEMC and EEMC system at the trigger and 
reconstruction level. A high-tower (HT1 / HT2) trigger is based on 
an energy threshold above 2.6 (3.5) GeV for a single tower ($\Delta \eta \times \Delta \phi = 0.05 \times 0.05$) 
whereas a jet-patch (JP1 / JP2) trigger is based on an energy
threshold above 4.5 (6.5) GeV for a group of towers over a region in $\eta$ and $\phi$ of $\Delta \eta \times \Delta \phi = 1.0 \times 1.0$.
Both triggers are taken in coincidence with a minimum-bias condition using the STAR Beam-Beam Counter. 

\begin{figure}[t]
\begin{minipage}[t]{5cm}
\includegraphics[height=5.0cm,width=10.0cm]{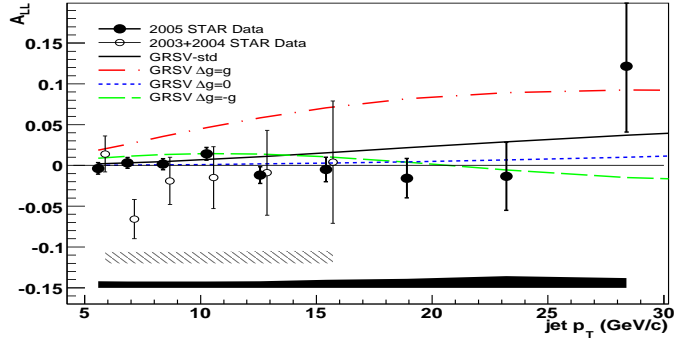}
\end{minipage}
\hfill
\begin{minipage}[t]{5.0cm}
\parbox[b]{5cm}{\caption{\it Longitudinal double-spin asymmetry $A_{LL}$ for inclusive jet production production at $\sqrt{s}=200\,$GeV as a function of $p_{T}$ ($0.2 < \eta < 0.8$) 
for Run 5 in comparison to several gluon polarization scenarios.}}
\end{minipage}
\hfill
\label{fig:all_2005_jets}
\end{figure}

Figure \ref{fig:all_2005_jets} shows the measured longitudinal double-spin asymmetry $A_{LL}$ for neutral pion production as a function
of $p_{T}$ together with different gluon polarization scenarios as described above. The error bars include statistical uncertainties only.
The systematic error band includes contributions from the neutral pion yield extraction and background subtraction, remaining background,
possible non-longitudinal spin contributions and the relative luminosity uncertainty. This analysis is based on a fraction of the Run 5
data sample for a restricted pseudo-rapidity region. Data from Run 6 will include the full acceptance of the STAR BEMC. 
The data tends to disfavor a large positive gluon polarization scenario. 

In addition to the first neutral pion analysis of $A_{LL}$ at mid-rapidity, the STAR collaboration has recently presented
the first preliminary result of neutral pion production using the STAR EEMC in the pseudo-rapidity acceptance
region of $1.086<\eta<2.0$ from Run 5 \cite{ref_spin2006_jason}. The data 
are consistent with zero over the $p_{T}$ range of $3<p_{T}<7\,$GeV/c, 
albeit with large statistical uncertainties.
The systematic uncertainties are comparable to the statistical uncertainties and dominated by beam induced background. These background
contributions were observed to be suppressed by a factor 20 during Run 6 relative to those in Run 5 profiting from the installation of 
shielding to reduce beam induced background prior to Run 6. This EEMC based analysis provides an important baseline measurement for
future prompt photon measurements in the STAR EEMC acceptance region.

The STAR collaboration has presented the first measurement of the longitudinal double-spin asymmetry $A_{LL}$ for
inclusive charged pion production during Run 5. The asymmetries are
calculated over the transverse momentum region $2<p_{T}<10\,$GeV/c and compared to several gluon polarization scenarios as described earlier.
This analysis is unique in that the 
difference $A_{LL}(\pi^{+})$ - $A_{LL}(\pi^{-})$ tracks the sign of the gluon polarization, due to the opposite signs of 
the polarized distribution functions for up and down quarks. 
The STAR TPC offers robust reconstruction and identification of 
charged pions over the transverse momentum range $2<p_{T}<10$ GeV/c.  Particle identification in the TPC is accomplished 
using measurements of ionization energy loss of TPC hits.  Figure \ref{fig:all_2005_pi-_pi+} shows preliminary results during Run 5 for
charged pion production. The measured asymmetries are compared to theoretical predictions for $A_{LL}$ based on different gluon polarization 
scenarios. The fragmentation functions for $\pi^{+}$ and $\pi^{-}$ are based on the KKP fragmentation functions \cite{ref_kkp}. 
This first measurement of $A_{LL}$ for charged pion production disfavors as well a large gluon polarization scenario. 
Several systematic checks have been performed. The leading systematic uncertainty accounts for the bias introduced by the trigger used for
this analysis. This trigger is based on a jet patch trigger, which introduces a bias towards jets with a large fraction of neutral
energy. The impact of this trigger on the charge pion asymmetry analysis has been estimated using a MC sample and amounts to approximately 
$5\cdot 10^{-3}$, which is comparable to the statistical uncertainty of the first $p_{T}$ bin. In addition, asymmetries were calculated
for trigger-jets in comparison to away-side jets. Both asymmetries have been found to be consistent within statistical uncertainties. 

The STAR collaboration has recently released preliminary results of the measurement of the longitudinal double-spin
asymmetry $A_{LL}$ for inclusive jet production, which is shown in Figure \ref{fig:all_2005_jets}
as a function of $p_{T}$ for $5<p_{T}<30\,$GeV/c in comparison to several gluon polarization scenarios as described earlier 
\cite{ref_spin2006_dave}. 
Jets are reconstructed using a midpoint cone clustering algorithm using a cone radius of $0.4$. This algorithm is fed in the case of STAR
by reconstructed electromagnetic clusters in the STAR BEMC and tracks from the STAR TPC. 
The uncertainties show statistical uncertainties only. Various systematic effects have been studied. The dominant contribution
to the systematic uncertainties are due to false asymmetries, trigger bias and jet reconstruction bias. The 2005 $A_{LL}$ inclusive
jet measurement is found to be in good agreement with the previous 2003/2004 measurement. The current measurement extends the
$p_{T}$ region to larger values, where the quark-gluon contribution starts to become important. This analysis rules out a large
gluon polarization scenario. It provides the most precise $A_{LL}$ measurement to constrain the gluon polarization
of the proton to date at RHIC. 

Taking all current PHENIX \cite{ref_phenix_overview1, ref_phenix_overview2} and STAR $A_{LL}$ measurements together 
in comparison to different NLO perturbative QCD predictions for $A_{LL}$ yields a consistent picture which rules out a large gluon polarization scenario. 
The STAR inclusive measurements will benefit enormously from the increased data sample in Run 6, the larger beam polarization and
the wider detector acceptance at central rapidity with the completion of the STAR BEMC.
A critical aspect to extract the gluon polarization of the proton is to perform a global analysis of several $A_{LL}$ measurements obtained
from the PHENIX \cite{ref_phenix_overview1, ref_phenix_overview2} and STAR collaborations taking into account a constraint of polarized parton distribution 
functions at high Bjorken-x
values by several polarized fixed-target DIS experiments. All the required full NLO calculations exist and have been incorporated
into a full NLO global analysis framework \cite{ref_gs, ref_df, ref_marco_dis2006}. 

STAR will extend its existing inclusive jet measurements to di-jet measurements. This will allow a better constrain of the
underlying event kinematics to extract the shape of the gluon polarization in a global analysis \cite{ref_marco_dis2006}. 
Photon-jet coincidence measurements
are expected to provide a theoretically clean way to extract the polarized gluon distribution. Measurements at 
both $\sqrt{s}=200\,$GeV and $\sqrt{s}=500\,$GeV are preferred to maximize the kinematic region in $x$ as well as to provide a 
means to measure the effect of scaling violations at fixed Bjorken-x, but different $p_{T}$ values. 

The STAR collaboration has presented a proof-of-principle measurement of the longitudinal spin transfer 
$D_{LL}$ in inclusive $\Lambda$ ($\Lambda\rightarrow p \pi^{-}$) and $\bar{\Lambda}$ ($\bar{\Lambda}\rightarrow \bar{p}\pi^{+}$) 
production in polarized proton-proton collisions at a center-of-mass 
energy of $\sqrt{s}=200\,$GeV at a mean transverse momentum $p_{T}$ of about $1.3\,$GeV/c and a longitudinal momentum fraction of 
$x_{F}=7.5 \times 10^{-3}$. 
The measurement of $D_{LL}$ for inclusive 
$\Lambda$ and $\bar{\Lambda}$ production may provide constraints on strange (anti) quark polarization \cite{ref_lambda} and can yield new insight into
polarized fragmentation functions \cite{ref_trans}. The extension of the $p_{T}$ region to large values is essential. 

The production of ${\rm W^{-(+)}}$ bosons provides an ideal tool to study the 
spin-flavor structure of the proton. ${\rm W^{-(+)}}$ bosons are produced in 
${\rm \bar{u} + d(u + \bar{d}}$) collisions and can be detected through their 
leptonic decay ~\cite{[bs8],[bs3]}. 
An upgrade of the STAR forward tracking system is currently in preparation to provide the required 
tracking precision for ${\rm W^{-(+)}}$  charge sign discrimination in the acceptance
region of the STAR EEMC in the electron (positron) decay mode \cite{ref_upgrade_surrow}. The W physics program requires RHIC 
running of longitudinal polarized proton beams at a center-of-mass energy of $\sqrt{s}=500\,$GeV at
high luminosity and polarization.


\begin{thebibliography}{9}   
\bibitem{ref_bunce}G. Bunce {\it et al}., \ARNPS 50, 525 (2000).
\bibitem{ref_cad}M. Bai, these proceedings.
\bibitem{ref_fpd_an}J. Adams {\it et al.} (STAR Collaboration), \PRL 92, 171801 (2004).
\bibitem{ref_all_0304}B.I. Abelev {\it et al.} (STAR Collaboration), \PRL 97, 252001 (2006).
\bibitem{ref_spin2006_jan}J. Balewski, these proceedings, hep-ex/0612036.
\bibitem{ref_spin2006_adam}A. Kocoloski, these proceedings, hep-ex/0612005.
\bibitem{ref_spin2006_larissa}L. Nogach, these proceedings, hep-ex/0612030.
\bibitem{ref_spin2006_dave}D. Relyea, these proceedings.
\bibitem{ref_spin2006_frank}F. Simon, these proceedings, hep-ex/0612004.
\bibitem{ref_spin2006_jason}J. Webb, these proceedings.
\bibitem{ref_spin2006_qinghua}Q. Xu, these proceedings, hep-ex/0612035.
\bibitem{ref_A.Ogawa}A. Ogawa, these proceedings.
\bibitem{ref_E704}D. Adams {\it et al.} (E704 Collaboration), \PLB 261, 201 (1991); \PLB 264, 462 (1991).
\bibitem{ref_fpd_cross}J. Adams {\it et al.} (STAR Collaboration), \PRL 97, 152302 (2006). 
\bibitem{ref_sivers}D. Sivers, \PRD 41, 83 (1990); \PRD 43, 261 (1991).
\bibitem{ref_collins}J.C. Collins, \NPB 396, 161 (1993).
\bibitem{ref_qui_sterman}J. Qiu and G. Sterman, \PRD 59, 014004 (1998).
\bibitem{ref_efremov}A.V. Efremov, V.M. Korotiyan and O.V. Teryaev, \PLB 348, 577 (1995).
\bibitem{ref_kouvaris}C. Kouvaris {\it et al.}, hep-ph/0609238.
\bibitem{ref_alesio}U. d'Alesio and F. Murgia, private communications.
\bibitem{ref_an_hermes} M. Diefenthaler {\it et al.} (HERMES Collaboration), Proceedings of the 13th International
Workshop on Deep Inelastic Scattering (DIS2005), Madison, April 2005, hep-ex/0507013.
\bibitem{ref_bv} D. Boer and W. Vogelsang, \PRD 69, 094025 (2004).
\bibitem{ref_v_y} W. Vogelsang and F. Yuan, \PRD 72, 054028 (2005).
\bibitem{ref_werner}W. Vogelsang, private communications.
\bibitem{ref_star}K.H. Ackermann {\it et al.} (STAR Collaboration), \NIMA 499, 624 (2003).
\bibitem{ref_grsv}B.J\"{a}ger, M. Stratmann and W. Vogelsang, \PRD 70, 034010 (2004).
\bibitem{ref_grsv_global_fit}M. Stratmann and W. Vogelsang, \PRD 70, 034010 (2004).
\bibitem{ref_frank_cipanp}F. Simon {\it et al.} (STAR Collaboration), Proceedings of the 
Conference on the Intersection of Particle and Nuclear Physics (CIPANP 2006), Puerto Rico, June 2006.
\bibitem{ref_kkp}B.A. Kniehl, G. Kramer and B. P\"{o}tter, \NPB 597, 337 (2001).
\bibitem{ref_phenix_overview1}K. Barish, these proceedings.
\bibitem{ref_phenix_overview2}A. Deshpande, these proceedings.
\bibitem{ref_gs}M. Stratmann and W. Vogelsang, \PRD 64, 114007 (2001).
\bibitem{ref_df}D. de Florian, G.A. Navarro and R. Sassot, \PRD 71, 094018 (2005).
\bibitem{ref_marco_dis2006}M. Stratmann, Proceedings of the 14th International
Workshop on Deep Inelastic Scattering (DIS2006), Tsukuba, April 2006.
\bibitem{ref_lambda}Q.H. Xu, Z.T. Liang and E. Sichtermann, \PRD 73, 077503 (2006).
\bibitem{ref_trans}D. de Florian, M. Stratmann and W. Vogelsang, \PRD 81, 530 (1998).
\bibitem{[bs8]} C. Bourrely and J. Soffer, \PLB B314, 132 (1993). 
\bibitem{[bs3]} B. Dressler et al., \EPJC 18, 719 (2001); B. Dressler 
{\it et al}., \EPJC 14, 147 (2000).
\bibitem{ref_upgrade_surrow}B. Surrow {\it et al.} (STAR Collaboration), \NIMB 241, 293 (2005).

\end{thebibliography}
\end{document}